\documentclass[reprint,groupedaddress,amsmath,amssymb,aps,prl]{revtex4-2}

\usepackage{graphicx}
\usepackage{soul,color}
\usepackage{float}
\usepackage{comment}
\usepackage{lineno}
\usepackage{printlen}
\usepackage[up,bf,raggedright]{titlesec}
\titleformat*{\section}{\large\bfseries}
\usepackage{natbib}
\usepackage{hyperref}
\hypersetup{colorlinks,linkcolor={magenta},citecolor={blue},urlcolor={red}}
\usepackage{romannum}

\begin{document}
\preprint{APS/123-QED}	
\title{Towards terahertz nanomechanics} 
\author{Jiacheng Xie$^{1}$, Weifeng Wu$^{2}$, Mohan Shen$^{1}$, Patrick Fay$^{2}$, Hong X. Tang$^{1,*}$}

\address{$^1$Department of Electrical Engineering, Yale University, New Haven, CT, USA}
\address{$^2$\makebox{Department of Electrical Engineering,~University of Notre Dame, Notre Dame, IN, USA}\\$^*$hong.tang@yale.edu}
	
\date{\today}
\begin{abstract}
Advancing electromechanical resonators towards terahertz frequencies opens vast bandwidths for phononic signal processing. In quantum phononics, mechanical resonators at these frequencies can remain in their quantum ground state even at kelvin temperatures, obviating the need for millikelvin cooling typically required for GHz resonators. However, electrical actuation and detection of mechanical motion at such high frequencies present significant challenges, primarily due to the need for device miniaturization to support acoustic waves with nanometer-scale wavelengths. One effective strategy is to aggressively thin down piezoelectric thin films, ideally to a thickness on the order of the acoustic wavelength, which is in the tens of nanometers. In this work, we aggressively reduce the thickness of lithium niobate from 300\,nm to 67\,nm through several stages, and fabricate suspended Lamb-wave resonators at each thickness level. These resonators achieve resonant frequencies as high as 220\,GHz, doubling the previous record and approaching the terahertz frequency threshold. While ultrathin films exhibit a clear advantage in frequency gains, they also experience increased acoustic losses. Our results suggest that future advances in terahertz nanomechanics will critically rely on mitigating surface defects in sub-100\,nm thin films.

\end{abstract}

\maketitle

\noindent
The terahertz (THz) frequency range ($\sim$\,$0.3$--$10$\,THz), which bridges conventional microwave and infrared optical domains, represents a new frontier in science and technology. Thanks to THz waves' non-ionizing nature, high sensitivity to molecular vibrational/rotational modes, and strong penetration to non-polar and non-conductive materials \cite{pawar2013terahertz,shen2022recent}, THz technologies are widely adopted in industrial and medical applications \cite{naftaly2019industrial,amini2021review}, as well as basic sciences such as astronomy, condensed matter physics, and biochemistry \cite{kawano2013terahertz,fedorov2020powerful,plusquellic2007applications}. However, the full potential of this regime still faces a bottleneck: the lack of reliable sources, detectors, and control components \cite{song2011present}.

Nanoelectromechanical systems represent a powerful platform to address this challenge. Capable of interchanging energy between electrical and mechanical domains, they are critical components for signal transduction and processing. They are also highly sensitive to mass, force, and temperature \cite{yang2006zeptogram,chaste2012nanomechanical,moser2013ultrasensitive,ricci2019accurate,zhang2013nanomechanical,laurent201812}. These properties make them ideal building blocks for THz detection systems that require compactness and high scalability compared to complex optical platforms involving pump-probe techniques \cite{juve2010probing}, and Raman and Brillouin scattering spectroscopy \cite{rozas2009lifetime}.

Beyond classical applications, nanomechanical resonators are also promising platforms for fundamental quantum studies \cite{rips2013quantum, stannigel2012optomechanical,wollack2022quantum,arndt2014testing,forstner2020nanomechanical,clerk2020hybrid,han2021microwave}. Ultrahigh-frequency THz resonators can provide fundamental insights into quantum behaviors at more accessible temperatures \cite{velez2019preparation}. For instance, a 300\,GHz resonator has a 99.2\% probability of being in the quantum ground state at 3\,kelvin, a temperature readily achievable by a cost-effective cryocooler, while GHz resonators require either cooling to stringent millikelvin temperatures using costly dilution refrigerators, or employment of sideband cooling techniques \cite{schliesser2008resolved} to achieve a similar thermal occupancy.

\begin{figure}[t]
\centering
\includegraphics[trim={0cm 0cm 0cm 0cm},clip,width=1\linewidth]{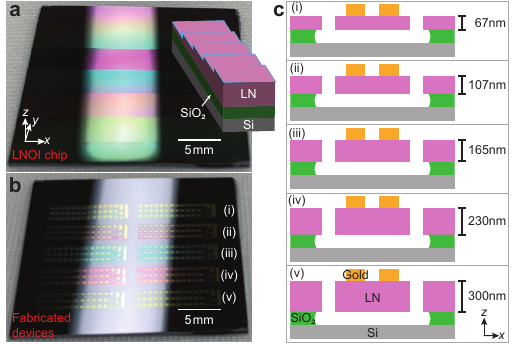}
\caption{{\bf Suspended Lamb-wave resonators (LWRs) with varying thicknesses.} {\bf a}, Image of the prepared lithium niobate-on-insulator (LNOI) chip featuring areas with multiple thickness levels: 67, 107, 165, 230, and 300\,nm. The variation in colors in the central region is due to the light reflection of a fluorescent lamp, highlighting the differing thickness levels across the chip. Inset shows the incremental thickness steps of lithium niobate (LN) film. {\bf b}, Image of the chip with fabricated devices. Five regions labeled (i) to (v) are marked. {\bf c}, Cross-sectional schematics of the suspended LWRs with varying LN thicknesses: 67\,nm (i), 107\,nm (ii), 165\,nm (iii), 230\,nm (iv), 300\,nm (v).}
\label{fig1}
\end{figure}

\begin{figure*}[t]
\centering
\includegraphics[trim={0cm 0cm 0cm 0cm},clip,width=1\linewidth]{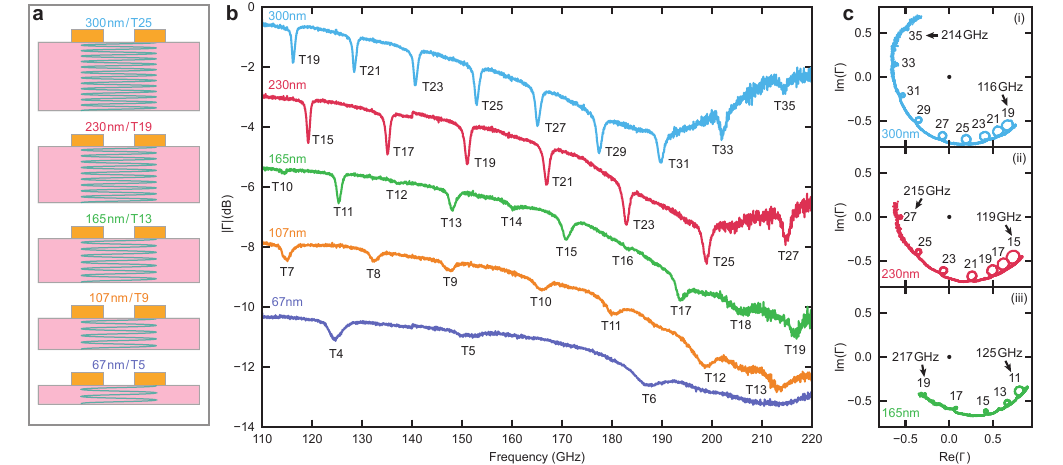}
\caption{{\bf Electrical responses of nanomechanical resonators.} {\bf a}, Illustration of thickness-shear (TS) mode displacement for varying film thicknesses, all with similar frequencies around 150\,GHz, with mode orders indicated. {\bf b}, Reflection ($\Gamma$) spectra for devices with thicknesses of 67\,nm, 107\,nm, 165\,nm, 230\,nm, and 300\,nm, plotted over 110-220 GHz. Mode orders are labeled adjacent to respective resonances. Curves are shifted vertically for clarity. {\bf c}, Smith chart representations of the spectra for devices with thicknesses of 300\,nm (i), 230\,nm (ii), and 165\,nm (iii). Mode orders are labeled adjacent to respective resonances. The origin is marked by a black dot.}
\label{fig2}
\end{figure*}

\begin{figure}[t]
\centering
\includegraphics[trim={0cm 0cm 0cm 0cm},clip,width=1\linewidth]{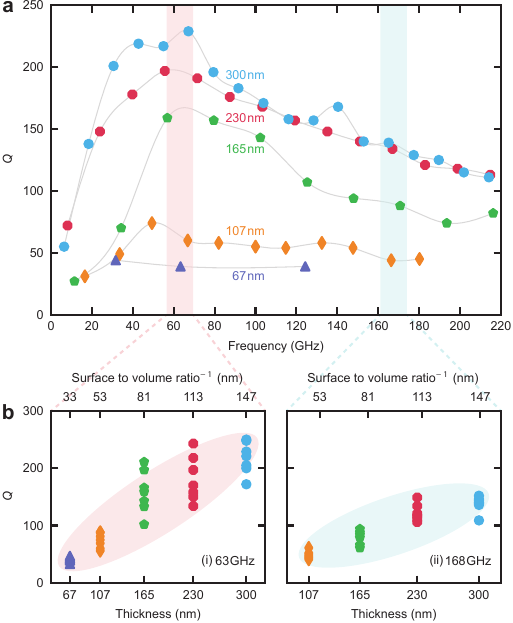}
\caption{{\bf Study of quality factors ($Q$s) in LN resonators.} {\bf a}, Extracted $Q$s of mechanical resonances across different mode orders for devices of varying thicknesses. Each curve represents a single device. {\bf b}, Scatter plots displaying $Q$s at approximately 63\,GHz (i) and 168\,GHz (ii) for devices of different thicknesses. Each color represents a group of multiple devices with the same thickness.}
\label{fig3}
\end{figure}

Recently, sub-THz electromechanical resonators operating near 110\,GHz have been demonstrated on thin-film lithium niobate (LN) platform \cite{xie2023sub,xie2024high,xie2024sub,kramer2025acoustic}, leveraging its excellent piezoelectric properties. However, advancing mechanical resonators towards the THz regime still presents significant challenges. At 300\,GHz, the acoustic wavelength of $z$-cut LN thickness-shear (TS) modes is around 10\,nm, which creates substantial difficulties in terms of efficient transduction. One effective strategy involves thinning down piezoelectric thin films to scale up the high-order thickness (T) modes, while still preserving electromechanical coupling even at hundreds of GHz. In this article, we systematically reduce the LN thickness from 300\,nm to 67\,nm to achieve multi-thickness levels. By fabricating suspended Lamb-wave resonators (LWRs) at each thickness level, we are able to study the relationship between mechanical quality factors ($Q$s) and thin film thicknesses. We demonstrate resonant frequencies up to 220\,GHz, approaching the THz frequency threshold.  However, thinner sub-100\,nm films show increased acoustic losses likely due to surface defects. We expect that future efforts to minimize these surface defects could lead to greater performance improvements in THz nanomechanics.\\

\noindent{\bf \large{Results}}\\
\noindent{\bf \normalsize{220\,GHz nanomechanics}}\\
Fig.\,\ref{fig1}a displays the optical image of lithium niobate-on-insulator (LNOI) chip as prepared, featuring incremental thickness steps: 67, 107, 165, 230, and 300\,nm (Methods). Fig.\,\ref{fig1}b illustrates the fabricated chip, consisting of LWRs across different thickness levels (Methods). In these images, optical interference leads to the perceived color difference among the films with different thicknesses. Fig.\,\ref{fig1}c shows the cross-sectional schematics of the LWRs, following the design in Ref.\,\cite{xie2023sub}. In this configuration, the resonators are suspended to minimize acoustic losses through the substrate. 

The electric field between the electrodes is mostly in the horizontal direction, enabling efficient excitation of the TS modes in suspended $z$-cut LN resonators via the large piezoelectric coupling element e$_{51}$. The TS mode displacement schematics of varying film thicknesses are shown in Fig.\,\ref{fig2}a. Clearly, to achieve a frequency around 150\,GHz, thicker films necessitate an increased mode order $n$. This escalation in mode order is accompanied by a reduction in the electromechanical coupling coefficient $K^2$, which is inversely proportional to the square of the mode order ($\propto 1/n^2$). Fig.\,\ref{fig2}b displays the reflection ($\Gamma$) spectra of devices, each with a length of 110\,\textmu m and varying thicknesses of 67, 107, 165, 230, and 300\,nm. The data are collected using a vector network analyzer with frequency extender modules for the D-band and G-band (Methods). Here, the spectral data are concatenated at 140\,GHz. Precise calibration methods are employed to minimize the stitching of spectra across different bands. Mechanical resonances around 220\,GHz are observed in Fig.\,\ref{fig2}b, including T35, T27, T19, and T13 modes for films of 300\,nm, 230\,nm, 165\,nm, and 107\,nm thicknesses, respectively. Additionally, the broadband tank resonance \cite{xie2023sub} shifts rightward with decreasing film thickness due to a reduced effective dielectric constant. Typically, even resonances are not effectively excited because the overlap between their strain fields and the applied electric field is mostly canceled out. However, we observe an intriguing phenomenon where, as film thickness decreases, resonances with even mode numbers begin to appear. For instance, in the 165-nm-thick device, modes T10, T12, T14, T16, and T18 can be observed, although their responses are significantly weaker than those of the odd resonances. As the LN thickness decreases to 107\,nm, the extinctions of the even resonances approach those of the odd resonances. In the case of the 67-nm-thick film, the odd resonances diminish significantly, while even resonances dominate, likely due to the increased surface damage, which inadvertently modifies the thin-film ferroelectric properties (supplementary document section \Romannum{2}). Additionally, to better visualize the complex impedance of nanomechanical resonances, we plot in Fig.\,\ref{fig2}c the Smith chart representations for films of 300\,nm, 230\,nm, and 165\,nm thickness, where the horizontal and vertical axes represent the real and imaginary parts of $\Gamma$, respectively. Even with lower mode orders, the resonance circles of the 165-nm-thick films are generally smaller than those of thicker films, indicating lower $Q$s for thinner films. This phenomenon is also evident in Fig.\,\ref{fig2}b, where thinner films exhibit wider mechanical linewidths.\\

\noindent{\bf \normalsize{Mechanical $Q$ and piezoelectric film thickness}}\\
The fitted intrinsic mechanical $Q$s (with a Lorentzian model) for various mode orders are shown as scattered points in Fig.\,\ref{fig3}, with each color indicating a different device thickness. Markers of the same color denote data from the same device shown in Fig.\,\ref{fig2}b. Due to the low signal-to-noise ratio of some resonances and uncertainty in fitting, we have omitted them from the plots. For relatively thick resonators, the $Q$s vary considerably with frequency, a phenomenon possibly linked to a frequency-dependent energy participation ratio of the lossy gold electrodes \cite{wang2024noncontact}. Further investigation is needed to better understand this phenomenon. In addition, we observe a clear general trend: resonators with greater thickness typically exhibit higher $Q$s compared to thinner ones. This is statistically evident in Fig.\,\ref{fig3}b, which compares the $Q$s of various devices with different thicknesses at approximately 63\,GHz (Fig.\,\ref{fig3}b(i)) and 168\,GHz (Fig.\,\ref{fig3}b(ii)). This negative correlation between $Q$ and the resonator's surface-to-volume ratio likely originates from surface-related losses, which are exacerbated by fabrication challenges on ultrathin films. At the 67\,nm scale, for instance, the film's structural fragility requires careful optimization of fabrication, including precise etch depth control and meticulous device handling. Recent studies have shown that fabrication processes such as ion milling, reactive ion etching, and thermal treatments can induce surface modifications of materials \cite{gruenke2025surface,soyer2005electrical,verbridge2006high,chen2009photonic,wang2017process,rodriguez2019direct}, which could lead to degraded piezoelectric performance in LN resonators. While the presence of this damaged layer has been observed via transmission electron microscopy \cite{gruenke2025surface}, the extent of the underlying damage and material impact remains to be investigated. To address this, we employ two complementary techniques: X-ray reflectometry (XRR) and selective wet etching of the damaged layer with hydrofluoric (HF) acid. These measurements provide quantitative insights into the thickness of the damaged layer, setting a practical lower bound on the usable thickness of fabricated resonators. 
\begin{figure}[]
\centering
\includegraphics[trim={0cm 0cm 0cm 0cm},clip,width=1\linewidth]{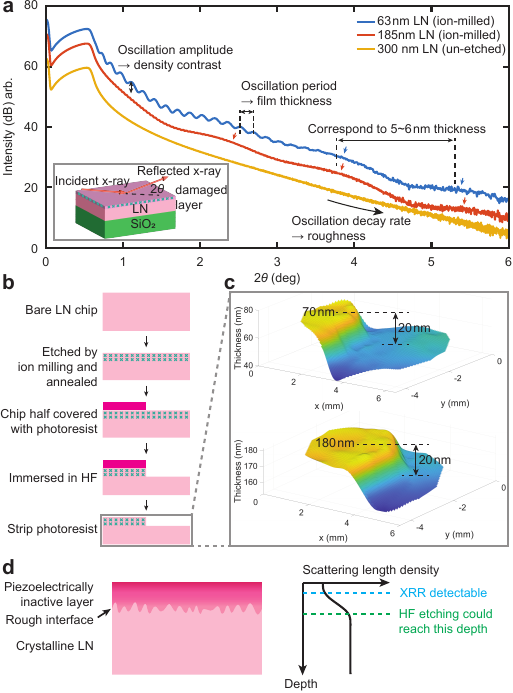}
\caption{{\bf Process-induced material damage.} {\bf a}, X-ray reflectometry (XRR) measurements of LN thin films of varying thicknesses (63, 185, and 300\,nm). Curves are shifted vertically for clarity. The 63\,nm and 185\,nm films are thinned from 300\,nm using ion milling. The inset illustrates the measurement schematic and the multilayer structure, including the hypothesized damaged layer. {\bf b}, Schematic illustration of the selective wet etching process. A bare LN chip is first thinned using ion milling, and annealed at 200\,$^\circ$C for 24\,h. Then half of the chip is masked with photoresist and immersed in 49\% hydrofluoric (HF) acid for 3\,minutes.
{\bf c}, Thickness maps obtained from optical interferometry for LN chips with initial thicknesses of 70\,nm and 180\,nm, respectively. 
A height difference of $\sim$20\,nm is observed between the HF-exposed and photoresist-protected regions, indicating the removal of a damaged surface layer. 
{\bf d}, Schematic illustration of the damaged surface layer. The scattering length density (SLD) gradually increases from the damaged (piezoelectrically inactive) surface to the crystalline LN. XRR primarily detects the upper region with a sharp SLD contrast (blue dashed line), while HF etching could remove chemically modified material deeper into the film (green dashed line).}
\label{fig4}
\end{figure}

\begin{figure*}[t]
\centering
\includegraphics[trim={0cm 0cm 0cm 0cm},clip,width=1\linewidth]{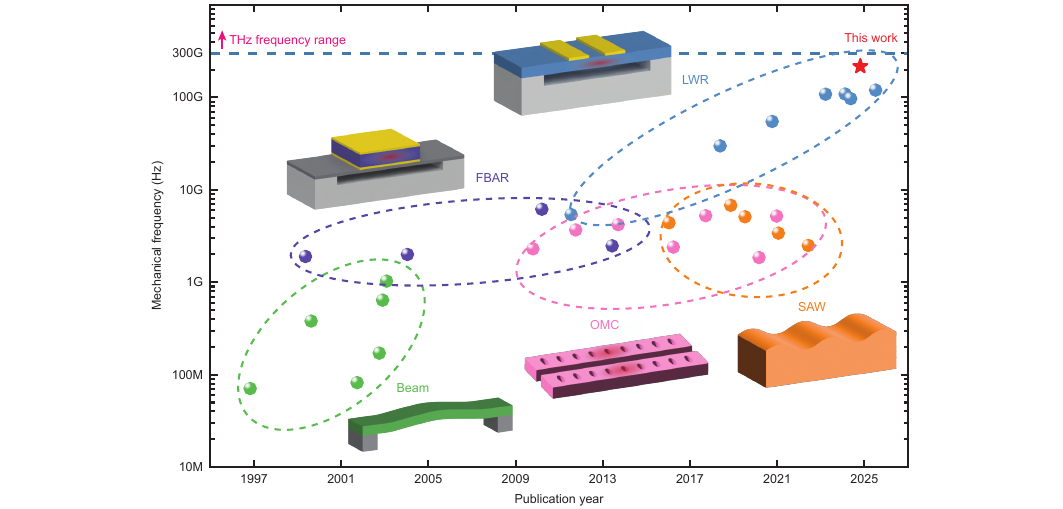}
\caption{{\bf Frequency scaling trend of micro/nanomechanical resonators} (published works are dated by their publication dates; this work is dated by its submission date). The mechanical platforms are categorized as follows: beam \cite{cleland1996fabrication,carr1999measurement,cleland2001single,sekaric2002nanomechanical1,sekaric2002nanomechanical2,henry2003nanodevice}, thin-film bulk acoustic resonator (FBAR) \cite{ruby1999pcs,gabl2004first,o2010quantum,kim2013detection}, surface acoustic wave (SAW) resonator \cite{manenti2016surface,fu2018high,shao2019phononic,mayor2021gigahertz,shao2022electrical}, optomechanical crystal (OMC) \cite{eichenfield2009optomechanical,chan2011laser,bochmann2013nanomechanical,balram2016coherent,hong2017hanbury,jiang2020efficient,mirhosseini2020superconducting}, Lamb-wave resonator (LWR) \cite{kadota20115,yang2018toward,yang202010,xie2023sub,xie2024high,xie2024sub,kramer2025acoustic}. The blue dashed reference line marks the THz frequency threshold (300\,GHz).}
\label{fig5}
\end{figure*}

XRR enables non-destructive probing of near-surface electron density profiles, making it well-suited to detect structural damages in thin films. We perform XRR on LN thin films of various thicknesses obtained via ion milling from 300\,nm. As shown in Fig.\,\ref{fig4}a, the main interference fringes corresponding to the full film thickness can be discerned for the 63- and 185-nm-thick samples. These oscillations arise from interference between X-rays reflected at the top and bottom interfaces of the film, and their spacing scales inversely with the film thickness. In the 300-nm-thick sample, the main oscillations are less visible because they are extremely closely spaced and difficult to resolve due to the large film thickness. The 5--6\,nm thickness inferred from XRR likely reflects only the portion of the surface damage with the most abrupt change in scattering length density (SLD). However, process-induced amorphization is typically more gradual. As a result, XRR may underestimate the total thickness of surface damage. To further investigate this possibility, we employ an alternative selective wet etching approach. In this experiment, we mask half of the processed chip with photoresist (S1813) and then immerse it in 49\% HF for 3 minutes, as illustrated in Fig.\,\ref{fig4}b. This photoresist layer protects the underlying surface from HF exposure. After stripping the resist, we measure the thickness profile using optical interferometry. As shown in Fig.\,\ref{fig4}c, a step height of $\sim$20\,nm is observed between the exposed and protected regions. Given that unprocessed LN is highly resistant to HF, this height difference indicates the removal of a damaged surface layer. This observation supports the hypothesis that the top $\sim$20\,nm of the film is modified by fabrication processes, resulting in degraded piezoelectric and mechanical performance in ultrathin devices. The discrepancy between these two methods suggests a graded surface modification profile, where the material transitions gradually from the piezoelectrically inactive surface to the crystalline interior, as shown in Fig.\,\ref{fig4}(d). 

In ultrathin films, such as the 67-nm-thick devices studied here, this damaged layer, together with surface roughness, can comprise up to $\sim$30\% of the total film thickness, greatly degrading mechanical performance. This offers a material-level explanation for the rapid decline in mechanical $Q$s observed as the film becomes thinner. As shown in Fig.\,\ref{fig3}b, extrapolation of the $Q$-thickness trend suggests that $Q$ continues to drop sharply with decreasing thickness, approaching values too low for practical electromechanical operation. Nonetheless, techniques such as chemical mechanical polishing (CMP) and atomic layer deposition (ALD), have the potential to overcome this challenge. CMP, a technique that combines chemical etching with mechanical polishing, is commonly reported to achieve ultrasmooth surface \cite{wu2006thinning,zhong2020recent}. On the other hand, ALD, with its precise thickness control to the nanometer range, can be employed to deposit ultrathin films with low defect densities \cite{zhang2011investigation}. Recently, it has been reported that ALD can be used to deposit sub-100\,nm-thick ferroelectric LN films \cite{ostreng2013atomic}. These efforts towards minimizing surface defects and improving the film quality of sub-100\,nm films are promising for unlocking significant potential and broad applications in THz nanomechanics.\\

\noindent{\bf \normalsize{Lamb-wave resonator's advantage for advancing THz nanomechanics}}\\
Over the years, a wide range of micromechanical and nanomechanical resonators, such as beam resonators, thin-film bulk acoustic resonators (FBAR), surface acoustic wave (SAW) resonators, optomechanical crystals (OMC), and Lamb-wave resonators, has been developed for applications in sensing, signal processing, communications, and quantum information studies. From a frequency scaling perspective, advances in fabrication techniques, material platforms, and design approaches have consistently driven mechanical frequencies upward. This progression in frequency capabilities is clearly demonstrated in Fig.\,\ref{fig5}, which charts the development from resonators operating below 1\,GHz two decades ago to this work with electromechanical resonators achieving a record-setting frequency of 220\,GHz. Unlike many resonator types whose frequencies are constrained by lithographically defined lateral dimensions, LWR frequencies depend primarily on film thickness. This relaxes the precision requirement in fabrication and enables frequency scaling by utilizing thinner films.\\

\noindent{\bf \large{Discussion}}\\
In this work, we demonstrate mechanical resonance frequencies up to 220\,GHz based on thin-film LN. By systematically thinning down LN through several stages and fabricating LWRs at each thickness level, we are able to study the correlation between mechanical $Q$s and resonator's surface-to-volume ratio. This paves the way for further optimizing sub-100\,nm film quality for improved device performance in THz nanomechanics. Additionally, we trace the evolution of frequency scaling in mechanical resonators, highlighting the advantages of LWRs in progressing towards THz frequencies. Our demonstration of 220\,GHz operation represents a critical advance toward the domain of THz nanomechanics, a regime with new physical challenges and application potential. At these frequencies, the acoustic wavelengths become comparable to sub-100\,nm thicknesses, making device performance increasingly sensitive to nanometer-scale imperfections and pushing both material characterization and fabrication to their limits. Operating at these high frequencies also offers an advantage for probing the Landau-Rumer regime, where phonon-phonon scattering behavior changes fundamentally and the $fQ$ product is predicted to scale favorably with frequency. Furthermore, by accessing a spectral region relevant for molecular fingerprinting, this advance lays the essential groundwork for integrated THz spectroscopy and sensing systems, and enables hybrid platforms with emerging THz photonics and electronics.\\

\noindent{\bf \large{Methods}}\\	
\noindent\textbf{Nanofabrication}\\
Initially, an LNOI chip with multiple thickness stages from 67 to 300\,nm is prepared. These different thicknesses are achieved through multiple blanket ion-milling cycles, with Si chips placed on top of the completed LN regions between cycles to prevent further etching. For device fabrication, a gold electrode pattern is defined through electron-beam lithography (EBL), with polymethyl methacrylate (PMMA) as the resist, followed by a liftoff procedure. Next, the release window pattern is defined using hydrogen silsesquioxane (HSQ) resist and then argon-ion-milled to fully remove the LN layer within the designated area. Finally, the mechanical resonators are suspended by etching away the silicon dioxide beneath the LN beam with buffered oxide etchant (BOE). A detailed fabrication process flow is illustrated in the supplementary document section \Romannum{1}.
	
\noindent\textbf{Calibration}\\
Scattering parameters are measured using a Keysight E8361C network analyzer with frequency extenders covering 67-110\,GHz (N5260-60003), the D-band (WR6.5-VNAX), and the G-band (V05VNA2-T/R-A). For each band, a tier-1 calibration is initially performed using off-wafer calibration standards, applying the line-reflect-reflect-match (LRRM) method. Subsequently, a tier-2 calibration is performed with on-chip calibration standards, utilizing the through-reflect-line (TRL) calibration technique. TRL transmission line parameters: signal width 40\,\textmu m, ground width 135\,\textmu m, signal-ground gap 12.5\,\textmu m.
\bibliography{references}

\noindent{\bf Data availability}\\
\noindent The data that support the findings of this study are available from the corresponding authors upon reasonable request.

\vspace{8mm}\noindent{\bf Acknowledgements}\\
This project is supported by the Air Force Office of Sponsored Research (AFOSR MURI FA9550-23-1-0338) and in part by the Defense Advanced Research Projects Agency (DARPA OPTIM HR00112320023). The part of the research that involves lithium niobate thin film preparation is supported by the US Department of Energy Co-design Center for Quantum Advantage (C2QA) under Contract No. DE-SC0012704. This collaborative effort also benefits from the support of a joint NSF FUSE program under award number 2235377 (PF and WFW). The authors would like to thank Yong Sun, Lauren McCabe, Kelly Woods, Michael Rooks, and Min Li for their assistance in device fabrication and characterization. The fabrication and characterization of the devices were performed at the Yale School of Engineering \& Applied Science (SEAS) Cleanroom, the Yale Institute for Nanoscience and Quantum Engineering (YINQE), and the Yale West Campus Materials Characterization Core.

\vspace{1mm}\noindent{\bf Author contributions}\\
HXT and JCX conceived the idea. JCX designed the devices with the help of WFW. JCX fabricated the devices with the help of MHS. WFW, JCX, and PF carried out the device measurements. JCX processed the data and provided the analysis. JCX wrote the manuscript with input from all the authors. HXT supervised the project.

\vspace{1mm}\noindent{\bf Competing interests}\\
The authors declare no competing interests.

\end{document}